\documentclass[aps,twocolumn,floatfix, superscriptaddress]{revtex4}
\usepackage[utf8]{inputenc}
\usepackage[breaklinks, colorlinks=true, urlcolor=blue, anchorcolor=blue, citecolor=blue, filecolor=blue, linkcolor=blue, menucolor=blue, linktocpage=true, pdfproducer=medialab]{hyperref}
\usepackage{amssymb,amsmath,graphicx,times}
\usepackage{mathptmx}
\usepackage{bm}
\usepackage[usenames]{color}
  \expandafter\let\csname equation*\endcsname\relax
  \expandafter\let\csname endequation*\endcsname\relax
 \usepackage{dsfont}
\newcommand{\ket}[1]{\vert #1 \rangle}
\newcommand{\bra}[1]{\langle #1 \vert }
\newcommand{\mean}[1]{\langle #1 \rangle }
\def\lc{\left\lfloor}   
\def\rc{\right\rfloor}
\newcommand{\Tr}[1]{\text{Tr} \left[ #1 \right] }

\begin{document}

\title{Bipartite unitary gates and  billiard dynamics in the Weyl chamber}

\author{Antonio Mandarino}
\email{mandarino@cft.edu.pl}
\affiliation{Center for Theoretical Physics, Polish Academy of Sciences, Al. Lotnik\'{o}w 32/46, 02-668 Warszawa, Poland}
\author{Tomasz Linowski}
\affiliation{Faculty of Physics, University of Warsaw, 
ul. Pasteura 5, 02-093 Warszawa, Poland}
\author{Karol \.Zyczkowski}
\affiliation{Center for Theoretical Physics, Polish Academy of Sciences, Al. Lotnik\'{o}w 32/46, 02-668 Warszawa, Poland}
\affiliation{Institute of Physics, Jagiellonian University, ul. {\L}ojasiewicza 11, 30--348 Krak\'ow, Poland} 

\begin{abstract}
\noindent
Long time behavior of a unitary quantum gate $U$,
acting sequentially on two subsystems of dimension $N$ each, 
is investigated. We derive an expression describing an arbitrary iteration
of a two-qubit gate making use of a link 
to the dynamics of a free particle in a $3D$ billiard.
Due to ergodicity of such a dynamics an average along a 
trajectory $V^t$ stemming from a generic two-qubit gate $V$
in the canonical form tends for a large $t$ to the average over
an ensemble of random unitary gates distributed according 
to the flat measure in the Weyl chamber - the minimal
$3D$ set containing points from all orbits of locally equivalent gates.
Furthermore, we show that for 
a large dimension $N$ the mean entanglement 
entropy averaged along a generic trajectory 
coincides with the average over the ensemble of random unitary matrices
distributed according to the Haar measure on $U(N^2)$.
\end{abstract}

\maketitle
\section{Introduction}
Since the dawn of the quantum technologies era, understanding 
the role played by nonlocal quantum gates has attracted  
a vivid 
interest, for computational \cite{NC00}
or control \cite{DAL08} protocols.  
Nonlocal unitary gates can be described by evolution operators 
associated to an interaction Hamiltonian $H$, that couples both subsystems.
A lot of attention has been devoted to implement experimentally controlled
gates in different settings,
ranging from ions in optical traps to condensed matter systems  
\cite{ION, NMR, phot}.

Several quantities designed to characterize the nonlocality of 
quantum gates focused on two-qubit unitaries \cite{KBG01,KC01,HVC02,ZVWS03,WSB03,Ni+03}, 
and the mean ability to create entanglement \cite{ZZF00, Za01}. 
Nevertheless a deeper investigation on 
nonlocal quantum gates is keeping going on in several directions, e.g.    
dynamical production of quantum correlations  \cite{AV07, JMZL17, GPPZ13},
time-optimal evolution and quantum brachistochrone problem \cite{BRAC1, BRAC2, BRAC3}, 
simulation of a particular quantum gate \cite{ZZSB15, OZ17} and 
extension to gates coupling subsystems of larger dimensions \cite{GEO1, GEO2}. 

Usually in a quantum protocol scheme a time dependent Hamiltonian 
is required to efficiently generate a target unitary gate.
In this work we study time independent Hamiltonians
acting sequentially on a quantum system a 
discrete number of times, $U^t= \exp (- i H t)$,
and analyze whether one
can reach in this way any typical orbit of locally equivalent gates.

Concentrating first on a class of two-qubit unitary gates we show that 
the time evolution of the nonlocality of their powers 
can be related to a  motion inside  
a classical three dimensional billiard.
As such a dynamics is known to be ergodic, the time and space averages
do coincide. Stronger statements, valid for generic unitary 
gates of dimension $N^2$ are established asymptotically.
These general results are directly applicable in various physical 
problems related to quantum control
and quantum information processing.
Furthermore, they can be interesting from the point of view
of the theory of random matrices and free probability.

The paper is structured as follows: in Sec.\ref{sec2} 
we recall the definition of local equivalence of 
quantum gates acting on two separate subsystems and the 
Schmidt decomposition of a unitary operator and 
we outline the geometric representation 
of two-qubit unitary quantum gates. 
In Sec.\ref{sec3} we  address the dynamics of a two-qubit gates
inside the Weyl chamber making a link to the dynamics of a free 
particle inside a 3-D billiard. In Sec.\ref{sec4}  we exploit the ergodic behavior 
to address some statistical properties of quantity of interest for quantum information purposes. 
Furthermore, we  discuss the impact of local random unitaries on the degree of nonlocality. 
In Sec.\ref{sec5} we extend our results to typical gates of size $N \times N$,  
while  Sec.\ref{sec6} closes the paper and contains concluding remarks. 

\section{Nonlocal quantum gates.}
\label{sec2}
A bipartite unitary quantum gate $U$ 
is \emph{nonlocal} if it cannot be written 
as the tensor product of two unitary matrices each acting on two different subsystems, 
$U \ne U_A \otimes U_B $. Therefore the action of a nonlocal gate
on a given state may alter the degree of its entanglement.
Any two unitary matrices $U, V \in SU(N^2)$  acting 
in the Hilbert space $\mathcal{H}= \mathcal{H}_A \otimes \mathcal{H}_B $ 
with $d_A=d_B=N$ are \emph{locally equivalent}, 
$U~\underset{\text{loc}}{\sim}~V$, 
if there exist local operations, $W_A \otimes  W_B$ and $W_C \otimes  W_D$, 
such that $
U=(W_A\otimes W_B)V(W_C \otimes W_D),$ \cite{KBG01,KC01} .
Any bipartite gate of size $N^2$ can be characterized 
by its \emph{Schmidt vector} $\vec{\lambda}(U)$ 
determined by the operator Schmidt decomposition, 
$U= N \sum_{j=1}^{N^2} \sqrt{\lambda_j} A_j \otimes B_j$,
and normalized to unity
-- see \cite{Ni+03,MKZ13} and Appendix \ref{ApA}. 
To quantify the nonlocality of such a gate one often uses
the {\sl entanglement entropy}  $S(U)$, 
also called the \emph{Schmidt strength}, 
equal to the Shannon entropy of the Schmidt vector,
\begin{align}
\label{entropies}
S(U) = - \sum_{i=1}^{N^2} \lambda_i \ln( \lambda_i )
, \quad 
S_L(U) = 1 -  \sum_{i=1}^{N^2} \lambda_i^2.
\end{align}
The second quantity $S_L(U)$, the linear entanglement entropy
related to purity of $\vec{\lambda}(U)$ and to
the entangling power of a gate \cite{Za01}, is easier to compute.

Consider now the group of all two-qubit gates, $\text{SU}(4)$, 
and its subgroup of local gates $\text{SU}(2) \otimes \text{SU}(2)$.
Making use of the Cartan decomposition of the group
it is possible to show \cite{KBG01,KC01} that any matrix $U \in \text{SU}(4)$,
is locally equivalent to a matrix written in the 
canonical {\sl Cartan form},
\begin{equation} \label{cartan}
U \underset{\text{loc}} {\sim} V_{(\alpha_1,\alpha_2, \alpha_3)} 
= \exp \left( i \sum_{k=1}^{3} \alpha_k \sigma_k \otimes \sigma_k \right). 
\end{equation}
Here $\sigma_1,\sigma_2, \sigma_3$ denote three Pauli matrices, 
while the phases $\alpha_k \in [0, \pi/2)$
form a vector $\vec{\alpha}\equiv (\alpha_1,\alpha_2,\alpha_3)$ 
called \cite{HVC02} \emph{information content} of the gate. 
Note that the hermitian operator in the exponent in Eq.(\ref{cartan}) 
can be considered as the Heisenberg Hamiltonian of a spin chain 
with the nearest neighbor interaction \cite{AFOV08}. 

For any given unitary $U\in U(4)$ the canonical 
form $V$ is not uniquely defined, 
as there is a symmetry with respect to any element of the Weyl
reflection group. To override this redundancy one selects 
a suitably chosen set in the $\vec{\alpha}$--space \cite{HVC02,ZVWS03},
called  the \emph{Weyl chamber},
such that each point represents uniquely 
a local orbit in the space of two-qubit unitary gates.
For our analysis it will be convenient to
identify local orbits determined by vectors 
$(\alpha_1,\alpha_2,\alpha_3)$ and $(\alpha_1,\alpha_2,-\alpha_3)$
and to consider the tetrahedron $\Gamma$ in
the $\vec{\alpha}$--space determined by inequalities 
$\frac{\pi}{4}\geq \alpha_1 \geq \alpha_2\geq \alpha_3 \geq~0$, 
which forms a half of the Weyl chamber \cite{WVM+15}.

The number $m$ of the non-zero components of the information content $\vec{\alpha}$
allows us 
to distinguish three different classes of two-qubit gates. 
The first class $\Gamma_I,$ containing gates with $\vec{\alpha}= (\alpha_1,0,0) \equiv \alpha_{I}$, corresponds to an edge of the tetrahedron. 
The second class $\Gamma_{II},$ characterized by 
$\vec{\alpha}=(\alpha_1,\alpha_2,0)\equiv \alpha_{II}$, forms the lower face of the 
tetrahedron $\Gamma$.
The last class, $\Gamma_{III}$, contains generic vectors  
$\vec{\alpha} = (\alpha_1,\alpha_2,\alpha_3)\equiv \alpha_{III}$
and covers the full measure of the 3D set  $\Gamma$.

\section{Dynamics of two--qubit gates.}
\label{sec3}
Time evolution of a typical gate 
belonging to a selected class can be analyzed by
mapping the dynamics of the information content, $\vec{\alpha}(V^t)$, 
to the position of a free point particle moving in a bounded domain. 
The case  $\Gamma_I$ can be considered as a dynamical analogue 
of the 1D motion of 
a free particle  between two walls placed at $0$ and $\pi/4$,
where it bounces elastically.
Iterations of a typical gate $(V_{\alpha_{I}})^t$ 
correspond to the motion along the axis
$\alpha_1$ and its position 
in $\Gamma_I$ after $ t \in \mathbb{N} $ time steps reads 
\begin{equation}
\label{eq:trW}
f( t \alpha_1)= \frac{\pi}{2} \left| \frac{2 t \alpha_1}{\pi} - \lc \frac{2 t \alpha_1}{\pi} + \frac{1}{2} \rc \right|,
\end{equation} 
where $\lc x \rc$ is the floor function.
This dynamics corresponds to the rotation 
at a circle modulo $\pi/2$
by an initial angle $\alpha_1$.
The reflection of the trajectory representing $V^t$ is
a consequence of folding of the entire 3D space 
into the Weyl chamber and identifying its two halves.
The argument of the floor function accounts for the number of reflections.
Knowing the information content $\alpha$ of $V$ 
one writes a simple expression for the content of its  power,
$(V_{\alpha_{I}})^t~=~\exp \left( i f( t \alpha_1) \sigma_1 \otimes \sigma_1 \right).$

\begin{figure}[t]
\includegraphics[width=1 \columnwidth]{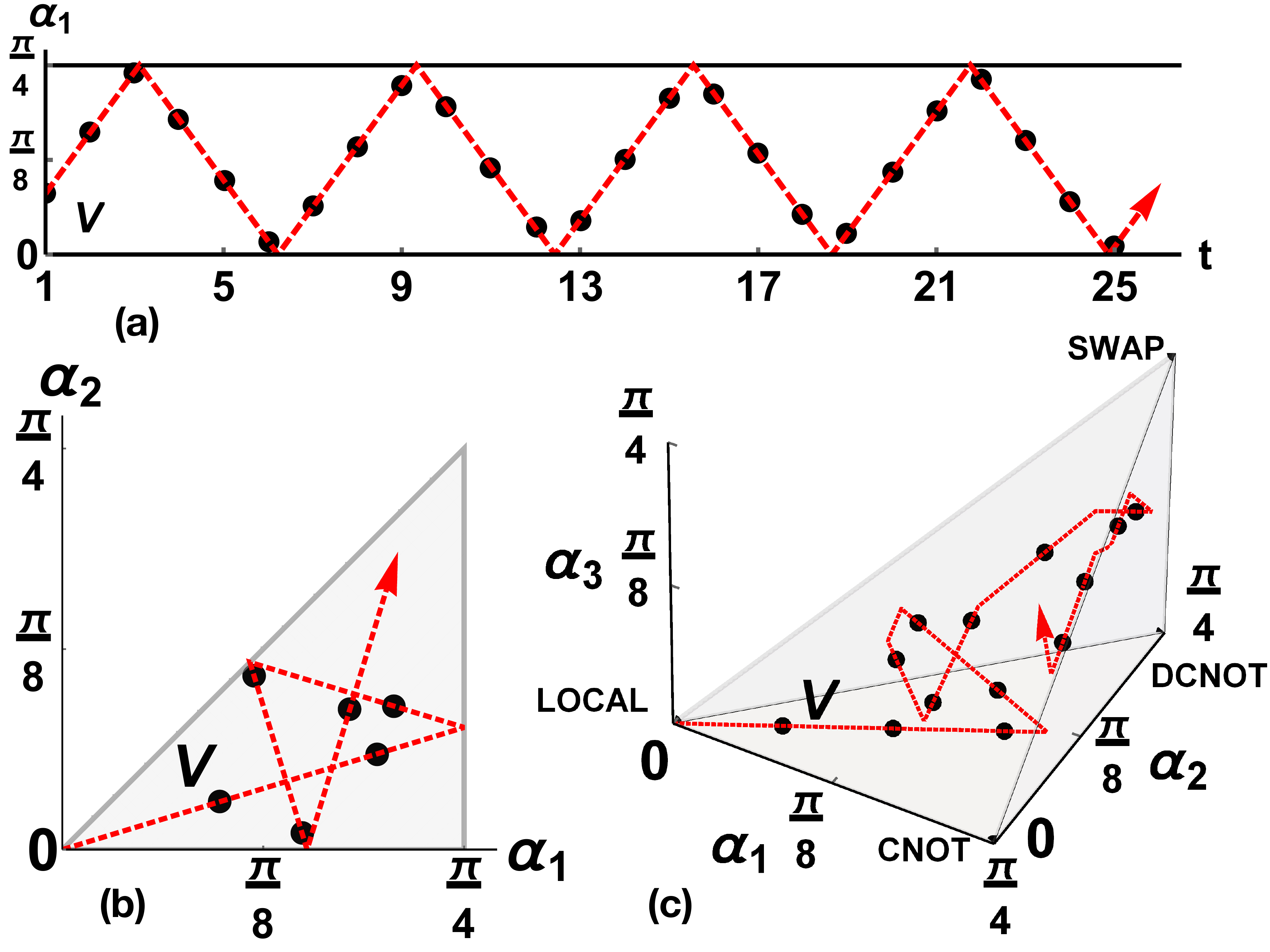} 
\caption{(a) Time evolution of $\alpha_1$ of a generic gate $V_{\alpha_{I}}^t$ 
after $t$ iterations according to Eq.(\ref{eq:trW}); (b) 
exemplary 2D trajectory of a typical gate:  $V_{\alpha_{II}}$ after 6 iterations; 
 (c) 3D trajectory in the tetrahedron $\Gamma$
generated by a typical gate $V$
in $\Gamma_{III}$ as prescribed by Eq.(\ref{eq:WCdyn}).
\label{fig:1}}
\end{figure}

Fig.~\ref{fig:1}(a) illustrates that the long time 1D dynamics of 
a typical gate $V\in \Gamma_I$ with a generic initial phase $\alpha_1$ 
explores entire accessible space.
This trajectory corresponds to the rotation of a circle by angle $\theta=\alpha_1$:
a well studied classical dynamical system \cite{MasurTab02}. 
If the rotation angle is not commensurate with $\pi$ 
the orbit of any initial point is dense, the dynamics is ergodic and 
the Lebesgue measure is invariant.
A gate like  $\sqrt{\text{CNOT}}$ corresponding to 
$\alpha_1=\pi/8$ is not generic, as it leads to a periodic orbit with period $4$.
If $\alpha_1$ is selected at random, with a flat measure in $[0,\pi/4]$,
the probability of such a case is equal zero.
The above argument can also be applied 
to the case of continuous time \cite{BS11,MKZ13},
which leads to the continuous flow $V(t)$.
For instance, $\vec{\alpha}(\sqrt{V}) = \frac12 \vec{\alpha}(V)$ 
corresponds to the position of the gate at $t=1/2$.

To generalize  Eq.(\ref{eq:trW}) for any initial conditions 
we introduce a function reordering the vector components in decreasing order, 
 ${\cal D}(\vec{x})= (x_1, x_2, x_3)$ 
such that $x_1 \geq x_2 \geq x_3$.
Then the information content of any power of a typical gate in $\Gamma_{III}$ reads 
\begin{equation}
\label{eq:WCdyn}
\vec{\alpha}(V^t) =   {\cal D} \bigl(
 f (t \alpha_1 ),f (t \alpha_2 ),f (t \alpha_3 ) \bigr),
\end{equation}  
see Appendix \ref{app1}. Exemplary 2D dynamics in $\Gamma_{II}$ is shown in Fig.\ref{fig:1}(b), 
where points corresponding to the gate $V^t$ 
are marked with black dots, while the dashed red lines describe 
the trajectory in the limit of continuous time. The accessible region in the
plane  $(\alpha_1,\alpha_2)$ forms a triangle.
We establish a link between the dynamics 
of a typical gate $V_{\alpha_{II}}$ and the motion of a 
classical particle within a right isosceles triangle billiard, 
that is known to be non-chaotic but ergodic \cite{Hurt97}.
In the 3D case $\Gamma_{III}$ a typical trajectory,
corresponding to a billiard inside the 
tetrahedron (such that its $48$ copies form a cube),
and sketched in Fig.~\ref{fig:1}(c),
explores uniformly the full volume of the set $\Gamma$.

\section{Entropy distribution for two--qubit gates}
\label{sec4}
Given the initial information content $\alpha$ of a gate 
we can describe the time evolution of entropies (\ref{entropies})
for three classes of gates $\Gamma_{m}$, with $m= I, II, III$. 
The entanglement entropy $S$ of a gate $V$ in the Cartan 
form (\ref{cartan})
as a function of the discrete time $t$ reads
\begin{equation}
\begin{split}
S(V_{\alpha_I}^t) &=- A_{1} \ln  A_{1} - B_{1} \ln B_{1}, \\
S(V_{\alpha_{II}}^t) &= -A_1  \left \{ A_2  \ln \left[ A_1  A_2  \right]+B_2  \ln \left[A_1  B_2 \right] \right\} 
 \\ &\quad  -B_1  \left \{ B_2  \ln \left[ B_1  B_2 \right] +A_2  \ln \left[ B_1  A_2 \right] \right \},
\end{split}
\label{eq:SvN}
\end{equation}
where using Eq.(\ref{eq:WCdyn}) we set 
$A_{k} = \cos^2 [f(t \alpha_k )], \   B_{k}=\sin^2[f(t \alpha_k )] , \ k=1, 2$. 
Analogous expression for $S(V_{\alpha_{III}}^t)$ can also be obtained but it is
too lengthy to be reproduced here.
Expressions for the linear entropy are simpler,
\begin{equation}
\begin{split}
S_L(V_{\alpha_I}^t) &=  \frac{1}{4} (1 -  c_1), \\
S_L(V_{\alpha_{II}}^t) &= 1- \frac{1}{16} \left( 3 + c_1 \right) \left ( 3 + c_2 \right ), \\
S_L(V_{\alpha_{III}}^t) &= \frac{9}{16}-\frac{1}{32} 
\big (4( c_1 + c_2 + c_3) +c^-_{12} + c^+_{12}+  2c_3(c_2 + c_1)\big), \\ 
\end{split}
\label{eq:SL}
\end{equation}
where $ c_i~=~\cos \left[ 4 f(t \alpha_i) \right], \
 c^{\pm}_{ij}~=~\cos \left[ 4 f(t \alpha_i) \pm 4 f(t \alpha_j)  \right], \\ \{ij\}=1,2,3 $.

We are going to analyze the distribution of entropies 
of powers of a gate in the canonical form (\ref{cartan}) 
with respect to averaging over time and over the parameter space. 
For a \textbf{fixed} initial gate $V_m$ 
with $m$ non-zero components
of the initial information content $\alpha$ we find the time evolution $V_m^t$,
compute the entropies, $S(V_m^t)$, and study the distribution $P(S)$.

For a generic choice of the point $\alpha$ 
the corresponding dynamics in the set $\Gamma_m$ is ergodic
and the Lebesgue measure $d^m \alpha$ is invariant.
Thus the time average obtained by integrating along a trajectory 
stemming from a typical gate $V_m$ 
and the average over the corresponding set $\Gamma_m$
in the parameter space with respect to the flat measure are equal.
In particular, the moments of the distributions of entropy
with respect to averaging over time and over parameter space do coincide,
\begin{equation}
\begin{split}
\overline{y^j}_{m,t} : = 
\lim_{T \rightarrow \infty }\frac{1}{T}\int_0^T y^j(V_{m}^t) dt = 
 \frac{1}{ \mathcal{N}_m }\int_{\Gamma_m} y^j(V) d^m \alpha.
= :\langle y^j \rangle_{m} . \\
\end{split}
\end{equation}
Here $y$ stands for one of the two entropies (\ref{entropies}) analyzed, 
integer $j$ labels moments of the distribution 
and the normalization factor 
$ \mathcal{N}_m = \frac{1}{m!} (\frac{\pi}{ 4} )^m$ is the volume of $\Gamma_m$.
Analytical values of first two moments of both
quantities, obtained by integration over 
subsets $\Gamma_m$ of the Weyl chamber, are collected in Table \ref{t:WeylChamber}. 
Looking at expressions (\ref{eq:SL}) for the linear entropy
we see that the oscillating terms do not contribute to the integral
and the average is determined by the constant terms only.

Note also that the average value of both entropies 
grows with the dimension $m$ of the subset 
$\Gamma_m$ of the Weyl chamber. 
This observation is consistent with a  surmise 
that a generic gate taken from the subset $\Gamma_m$ 
leads to a trajectory characterized by the average value 
of nonlocality which increases with the dimension $m$.

\begin{table}[t]
\caption{\label{t:WeylChamber} Mean values of the first two moments of 
Shannon entropy $S$ and linear entropy $S_L$ averaged over random
$2$-qubit gates from the $m$-dimensional subset $\Gamma_m$ of the Weyl chamber.} \vspace{2mm}
\begin{ruledtabular}
\begin{tabular}{c | c |  c | c } 
$m$  &  $1D$ \ set \ $\Gamma_I$ & $2D$  \ set \ $\Gamma_{II}$ & $3D$  \ set \ $\Gamma_{III}$ \\[1mm] \hline 
$\langle S \rangle$ & $ \ln 4 -1   \approx 0.386 $ & $ 2 (\ln 4  -1) \approx 0.772  $ & $ 1.028 $  \\[.8mm] \hline 
$\langle S_L \rangle$ & $1/4 = 0.25 $ & $ 7/16 = 0.4375 $ & $ 9/16 = 0.5625 $ \\[.8mm]  \hline
$\langle S^2 \rangle$ & $0.205$ & $ 0.709  $ & $ 1.143 $  \\[.8mm] \hline 
$\langle S_L^2 \rangle$ & $3/32 \approx 0.09 $ & $ 233/1024 \approx 0.228 $ & $ 351/1024 \approx 0.343 $ \\[.8mm]
\end{tabular}
\end{ruledtabular}
\end{table}

As the entropy $S$ belongs to a finite interval
equality of all the moments of two given distributions
implies that they coincide.
In this way we arrive at the following result.

{\bf Proposition 1.}
{\sl Probability distribution of the entanglement entropy $P_m^T(S)$
characterizing the sequence of gates $V_m^t$ generated from a given
two-qubit gate $V_m$ in the canonical form (\ref{cartan})
corresponding to a generic point $\vec{\alpha} \in \Gamma_m$
is equal to the entropy distribution  $P_m^{\Gamma}(S)$
obtained by averaging over the $m$--dimensional subset $\Gamma_m$ 
of the Weyl chamber with respect to the flat measure.}
 
\begin{figure}[t]
\includegraphics[width= 1 \columnwidth]{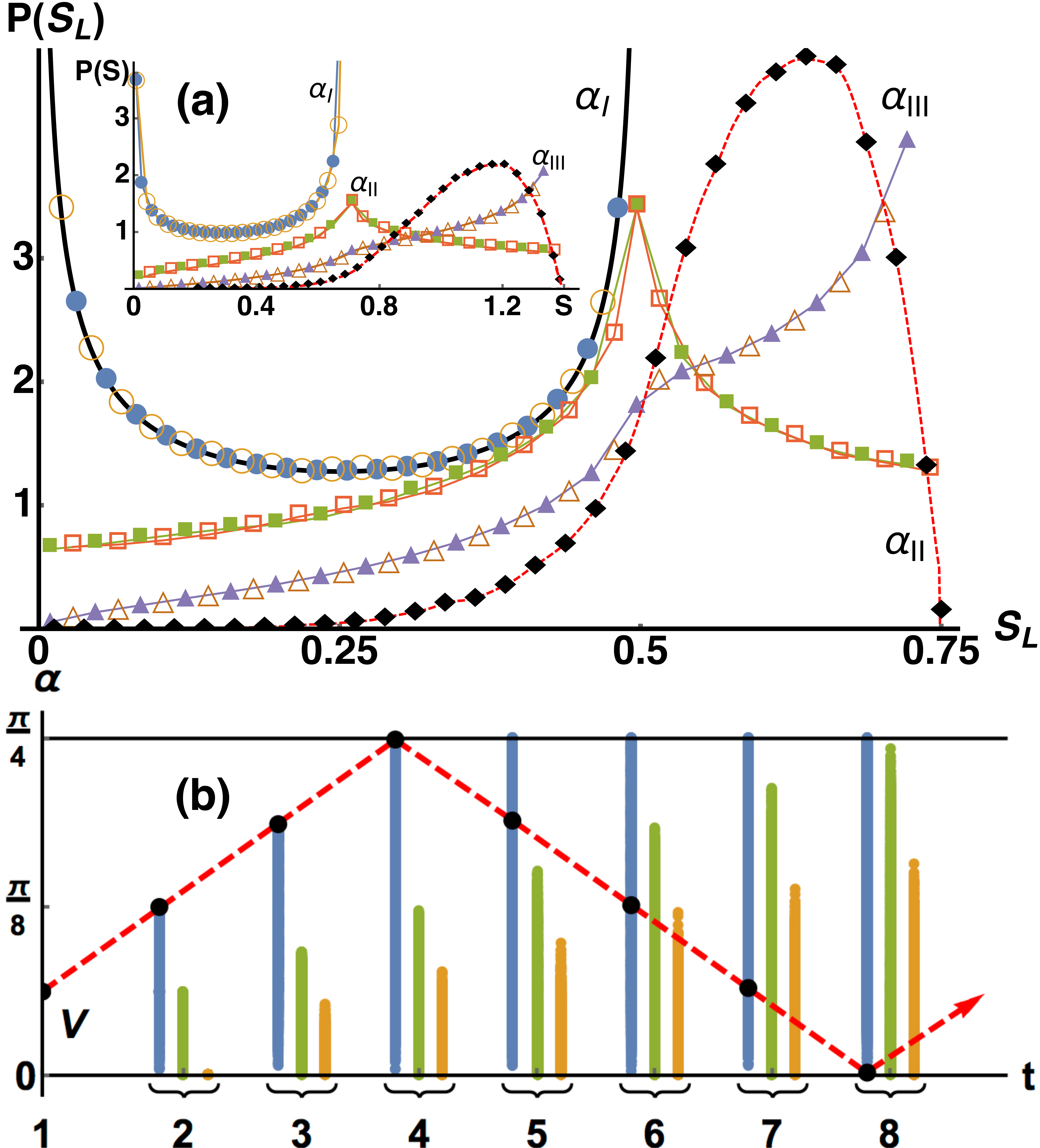}
\caption{\label{Fig:2} 
(a) Distribution of linear entropy $P(S_L)$
for canonical gates  $ V_{I} \in \Gamma_I$ (left curve), $ V_{II} \in \Gamma_{II}$ (middle curve) and
$ V_{III} \in \Gamma_{III}$ (right curve). 
Filled symbols in each plot refer to distribution of gates in 
the parameter space, while empty symbols to averaging in time 
over a single typical trajectory.
Solid black curve represents the arcsine distribution (\ref{asinlaw})
corresponding to the $1D$ case $\Gamma_I$.
Circles, squares and triangles refer to $m=I,II,III$,  respectively. 
For $m=II,III$ a line joining the squares or the triangles is plotted to guide the eye. 
The red dashed line and black diamonds represent 
the average over the Haar measure on $U(4)$.
Inset shows similar data concerning distributions of entanglement entropy $P(S)$.
(b) Time dependence of information content $\alpha$ for powers $V^t$ 
of a two-qubit gate $V$ in the canonical form (\ref{cartan})
and for an ensemble of random locally equivalent gates $U$.
Black dots correspond to $\alpha_1(V^t)$.
Blue, green and orange dots denote  $\alpha_m(U^t )$
for $m=1,2,3$, respectively. 
The red dashed line correspond to the time evolution of $\alpha_1(V^t)$ 
as in Fig.\ref{fig:1}(a)}
\end{figure}

An analogous statement holds also for the distributions of linear entropy $P(S_L)$
as visualized in Fig.\ref{Fig:2}(a). 
Filled symbols correspond to the distribution generated by 
an ensemble of $M=10^6$ random unitary matrices 
generated according to the Lebesgue measure in $\Gamma_{m}$. 
Empty symbols refer to distribution along a trajectory 
$V_{m}^t$ stemming from a single gate $V_{m}$ which
corresponds to a generic point $\alpha$ in $\Gamma_{m}$.
We emphasize that these distributions converge to the 
one related to averaging over $\Gamma_{m}$
independently on the choice of 
a typical starting point $V_m$.

Numerical results for two-qubit gates 
reported in Fig.\ref{Fig:2}a where we 
present a comparison of entropy distributions
referring to gates sampled randomly in the parameter space and 
to gates along the trajectory $V^t$ stemming from a typical gate $V$ written in the
canonical Cartan form.

Similar analysis can also be performed for any generalized  entropies 
of the normalized vector of Schmidt  coefficients  of any 
bipartite gate $U$, including all Renyi and Tsalis entropies \cite{BZ06}. 
However, in this work we concentrate 
on the linear entropy $S_L$, as this quantity is directly 
related  \cite{ZZF00} to entangling power of $U$.

One may pose a legitimate question, to what extend the results obtained
may depend on the choice of the initial gate $V$.
Wishing to study generic gates, which approximate
certain distinguished entangling gates,
we need to use a quantity describing the closeness of 
two selected gates. Following \cite{Fid}
we shall relay on the normalized Hilbert--Schmidt scalar product
of two matrices, also called the {\sl fidelity between two gates},
\begin{equation}
F(V_1,V_2)= \frac{1}{N^2}\left|\Tr {V_1^\dag V_2} \right|.  
\end{equation}

\begin{figure}[h!]
\includegraphics[width= 0.9 \columnwidth]{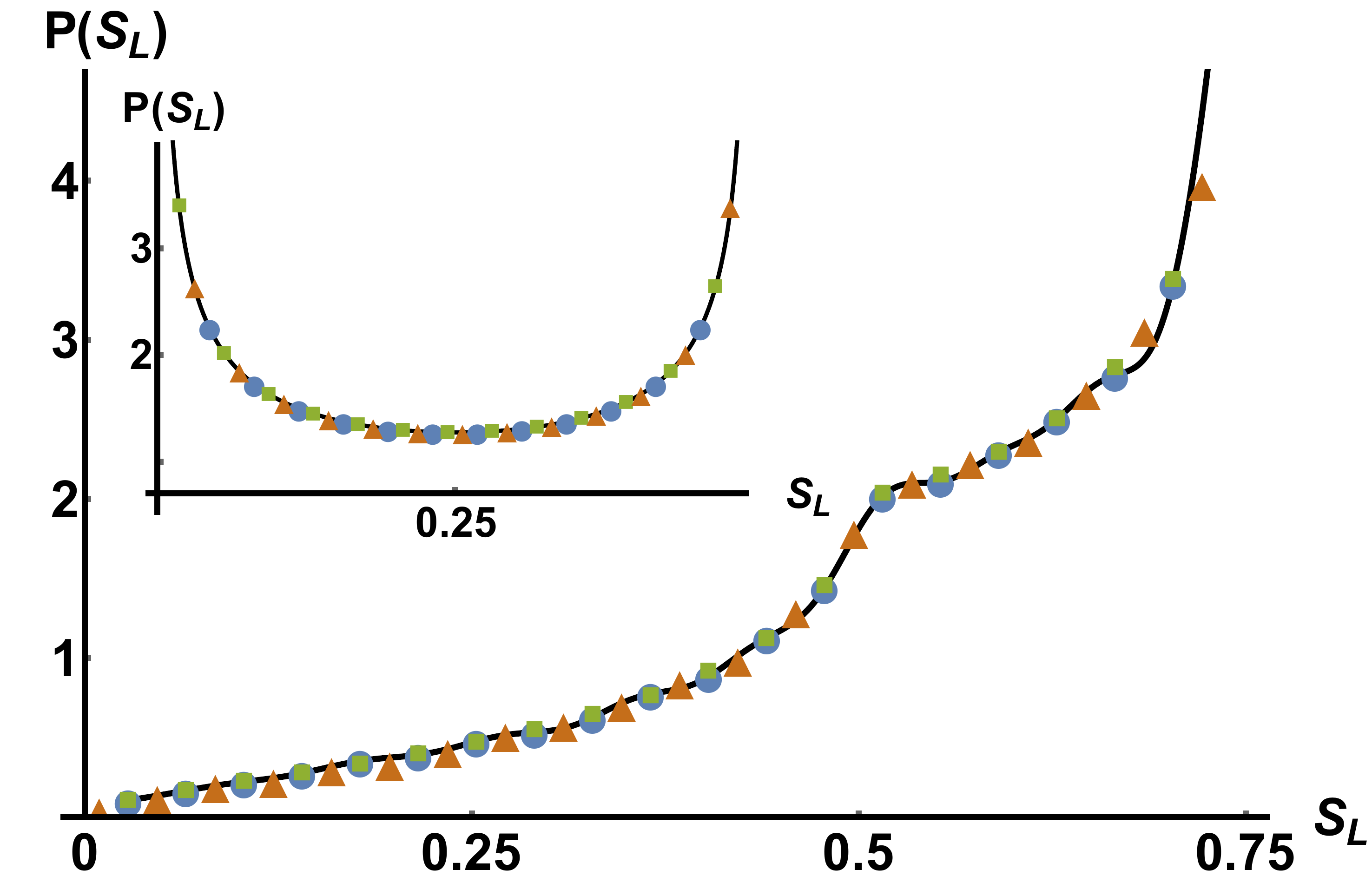}%
\caption{Distribution of linear entropy $P(S_L)$
averaged over a trajectory $V^t$ stemming from an initial gate
 $ V \in \Gamma_{III}$. 
Blue circles, green squares and yellow triangles refer to an initial gate $V$  
which approximates with fidelity $F \simeq 0.998$
a) local gate, b) $SWAP$ gate and c) $\sqrt{SWAP}$ gate.
Solid curve represents the inferred distribution. 
Inset shows similar data for initial gates $ V \in \Gamma_{I}$, 
which approximate with the same fidelity 
a) local unitary, b) $\sqrt{CNOT}$ and c) $CNOT$ gate
compared with the arcsine distribution represented by a solid line.
\label{fig:A2} } \end{figure}

In Fig.\ref{fig:A2} we demonstrate 
that the same asymptotic distributions are 
obtained for any typical initial gate. In particular we 
study the  probability distribution of linear entropy $S_L$ 
along the trajectories $V^t$ emerging from 
a typical gate $V$,  which up to a high fidelity approximate
selected gates including $CNOT$, $\sqrt{CNOT}$ and $\sqrt{SWAP}$.
Numerical data shown in the Figure 
confirm that the averaging along the trajectory
does not depend on the choice of the initial gate $V$,
provided it is taken as a typical gate in a 
given segment of the Weyl chamber $\Gamma$. 

Distribution averaged over the entire tetrahedron 
$\Gamma_{III}$ differs from the distribution over 
the ensemble of random unitary matrices,
which typically are not in the Cartan form, see Tab.\ref{t:N2} in Appendix \ref{ApC}.

To derive analytical form of the distribution of entropy 
we use the fact that the flat measure 
in the set $\Gamma_{m}$ of is invariant.
Taking the normalized measure $P(\alpha_1)= \frac{4}{\pi}$
in the simplest case, $m=1$,
inverting the first of Eq.(\ref{eq:SL}),  $ \alpha_1 = \frac14 \arccos(1 - 4 S_L)$, 
we obtain that $S_L$ is distributed in $ \Gamma_I$
according the {\sl arcsine law},
\begin{equation}
P(S_L) = \frac{1}{\pi \sqrt{S_L(\frac12 - S_L)}},
\label{asinlaw}
\end{equation}
which holds for $S_L\in[\frac12, 1]$.

The above distribution, up to a rescaling, 
gives the invariant measure of the logistic map 
in the regime of full chaos, $f(x)=4x(1-x)$. 
Analytical distributions for the other 
quantities can be obtained in a similar manner, 
but since their derivation is rather lengthy, 
it is omitted here. 


\subsection{Interlacing local unitary dynamics}
\label{subsec4a}
To demonstrate influence of local unitaries on the 
nonlocality of a sequence of bipartite gates \cite{JMZL17}
we compare the information content $\alpha(V^t)$
of iterated generic gate $V$ in the Cartan form
with the content $\alpha(U^t)$ of a locally equivalent gate $U=V Y_{\rm loc}$.
As the trajectory $\alpha(V^t)$ corresponds to a billiard 
dynamics in the tetrahedron, the evolution of $\alpha[(V Y_{\rm loc})^t]$
can be related to a dynamics of a billiard with a noise
which enhances the diffusion in the tetrahedron.
Fig.\ref{Fig:2}(b) shows dynamics of an ensemble of $10^4$ 
gates $U=V_I Y_{\rm loc}$, which differ from the gate $V_I\in \Gamma_{I}$
by a random local gate $Y_{\rm loc}$.
Due to interlacing local dynamics
the trajectory $U^t$ leaves the edge $\Gamma_{I}$
and the face $\Gamma_{II}$
and we have a strong numerical evidence that 
it eventually explores the entire $3D$ set $\Gamma_{III}$.

As shown in Fig.\ref{Fig:2}(b) also the second and the third component 
of the information content of $\alpha(U^t)$ 
become non-zero at $t=2$ and $t=3$ respectively. 
Therefore, sequential iterations of the analyzed gate U, 
interlaced by a generic local dynamics, 
allows the trajectory to explore full measure 
of the Weyl Chamber $\Gamma$.  


\section{Unitary gates of size $N x N$}
\label{sec5}
Consider a more general case of bipartite gates acting on two 
subsystems with $N$ levels each, 
described by unitary matrices from $SU(N^2)$.
Although the Cartan decomposition  
of this group exists for any dimension,
the adjoint action of $SU(N)$ forms for $N>2$ a small
subgroup of the orthogonal group $SO(N^2-1)$.
Hence using local unitary rotations from $SU(N)\otimes SU(N)$
 it is not possible for $N\ge 3$ to bring a generic unitary 
gate $U\in SU(N^2)$ to the diagonal form analogous to (\ref{cartan})
and to use the notion of information content \cite{HVC02,MKZ13}.
To describe nonlocality of such a gate we shall
then apply its Schmidt vector $\vec{\lambda(U)}$
and analyze both entropies (\ref{entropies}).

Selecting an initial gate $U_0$ 
we compute an average along the trajectory of length $T$, writing
$\overline{y} = \frac{1}{T} \sum_{t=1}^{T} y(U_0^t)$,
and study probability distributions $P_T(y)$.
Here $y$ represents one of the two entropies investigated.

Time averages are compared with mean values over 
circular ensembles of Dyson \cite{Dyson62}.
Asymptotic averages of both entropies with respect to 
Circular Unitary Ensemble (CUE) \cite{Metha}, equivalent to the Haar measure,
are \cite{ZZF00,MKZ13}
\begin{equation}
\label{eq:S_CUE}
\mean{S}_{U} = \ln N^2 - \frac12, \quad 
\langle S_L  \rangle_{U}  = \frac{N^2 - 1}{N^2 + 1}.
\end{equation}
Analogous averages for Circular Poissonian Ensemble (CPE),
which contains diagonal random unitary matrices with independent phases,
read \cite{LPZ14}

\begin{equation}
\label{eq:S_CPE}
\begin{split}
\mean{S}_{P} =& 1 - N + (N-1)^2 \ln \left[ N \left( \frac{N}{N-1} \right) \right], \\
\mean{S_L}_{P} =& \frac{( N - 1)^2}{N^2 }.\\
\end{split}
\end{equation}

\begin{figure}[t]
\includegraphics[width= 1  \columnwidth]{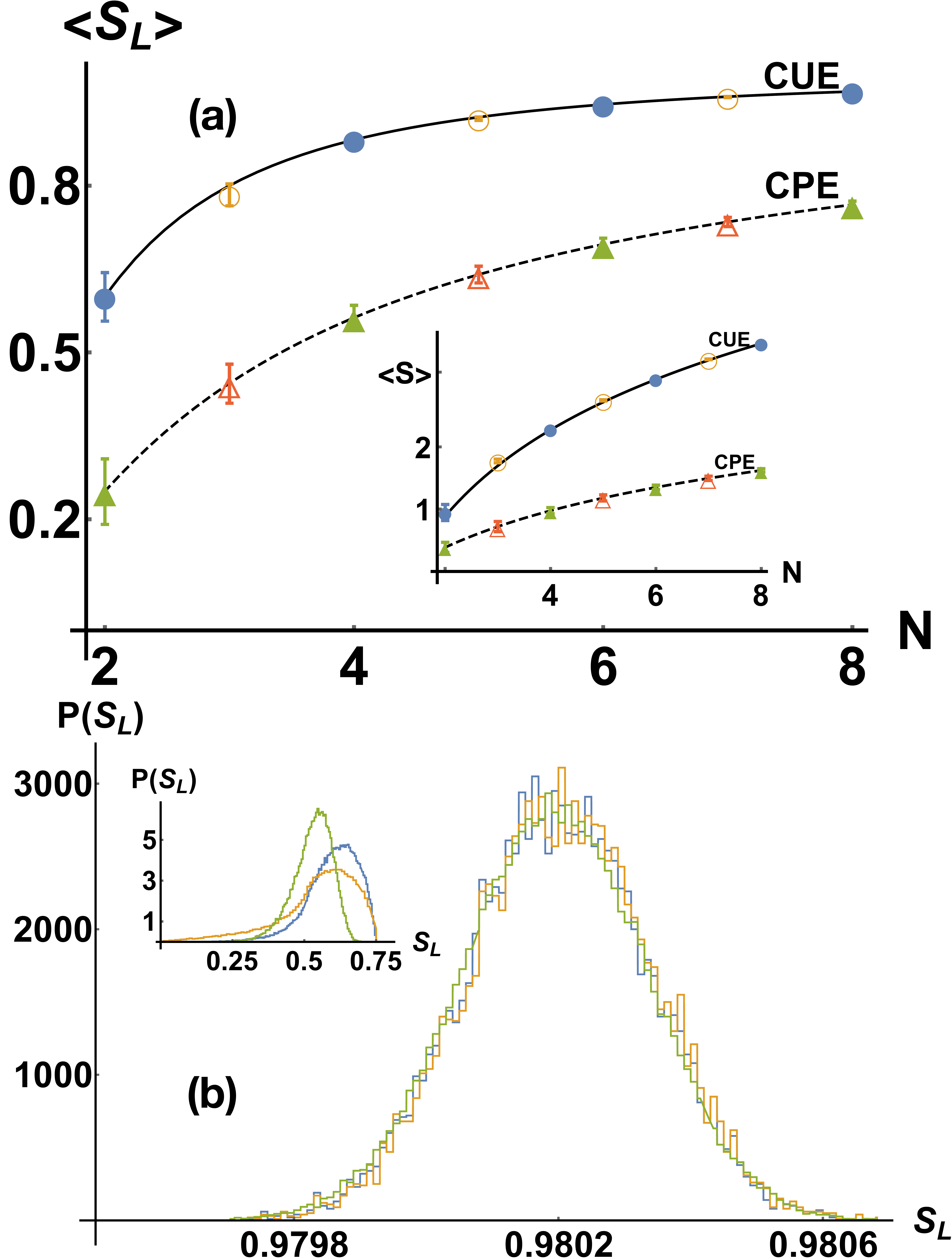}
\caption{(a) Average linear entropy $S_L(U)$ of random gates of size $N^2$.
Space averages  $\mean{y}$ denoted by filled symbols 
coincide with time averages $\overline{y}$ (empty symbols).
Solid curves represent interpolation between mean values (\ref{eq:S_CUE})
over CUE,  ($\circ$) denote averaging over trajectories 
stemming from generic $U$,  while dashed lines 
and triangles correspond to mean values (\ref{eq:S_CPE})
over CPE and over iterated generic diagonal gates.
Inset shows averages of Shannon entropy $S(U)$.
(b) Distribution $P(S_L)$ of a random samples of $M=10^6$ matrices: 
generated according to CUE (yellow), time trajectory initiated from a single CUE-matrix (blue) 
and characterizing the spectra of Wishart random matrices (green) 
practically coincide already for matrices of size $N^2=100$.
The differences between analogous distributions
obtained for $N^2=4$ are shown in inset with same notation.
\label{Fig:3} }
\end{figure}

Both entropies (\ref{entropies}) of Schmidt vectors of bipartite gates 
averaged numerically over a single trajectory 
are equal to averages over the corresponding ensemble
as demonstrated in Fig.~\ref{Fig:3}(a).
Blue filled circles are obtained by computing the mean values 
over a sample of $M=10^6$ random matrices $U$ distributed according 
to the Haar measure of the group $SU(N^2)$, while the yellow empty circles are related to the averages over
the trajectory of a fixed generic gate, $U_0$. 
Green filled triangles refer to averages over CPE
and empty orange triangles represent time averages over 
a single trajectory stemming from a random diagonal gate.

Let us now proceed to the distribution of entropy.
In the case of two-qubit diagonal gates of CPE 
the information content $\alpha$ has a single component,
so proceeding as in the case $\Gamma_I$
we infer that the distribution of linear entropy, $P(S_L)$,
is given by the arcsine law (\ref{asinlaw}).
Other results are established in larger dimensions. 
The level statistics of any diagonal $CPE_{N^2}$ matrix is Poissonian, 
and any power of a 
diagonal matrix is diagonal and has independent eigenvalues. This implies

{\bf Proposition 2.}
{\sl  The distribution of entropy along a trajectory $(U_{\rm diag})^t$ initiated 
from a generic diagonal gate is equal to the probability distribution 
over the circular Poissonian ensemble, 
\begin{equation}
 P_T(y)= P_{CPE} (y),
\end{equation}
where $y$ is one of the two entropies considered.} 

In contrast, for Haar random gates 
the distribution of linear entropy $S_L$ with respect to a given trajectory 
stemming from a random matrix $U_0$ coincides with 
the distribution over the ensemble in the asymptotic limit only -- see Fig.\ref{Fig:3}(b).
To evaluate the entropy we find the Schmidt vector $\lambda$
of a bipartite unitary matrix 
$U$ of size $N^2$ determined by the singular values (SV)
of the reshuffled matrix $U^R$ with entries 
$U_{\stackrel{\scriptstyle m \mu}{n\nu}}^R =U_{\stackrel{\scriptstyle m n}{\mu \nu}}$, 
since  $\sqrt{\lambda(U)}=SV(U^R)/N$ -- see Appendix \ref{ApA}.

For a large dimension $N$ the constraints imposed by the unitarity of $U$
become weaker, so that the spectral density of squared 
singular values of $U^R$ converges to the Marchenko-Pastur distribution,
characteristic to Wishart random matrices  $W=XX^{\dagger}$.
Here $X$ represents a random non-hermitian matrix 
of the Ginibre ensemble \cite{Metha}. Furthermore, the
distribution of entropy $P(S_L)$ for an ensemble of independent
Wishart matrices describes analogous distribution 
corresponding to the spectra of $Z^R (Z^R)^\dagger$,
although unitary matrices forming a trajectory, $Z=U_0^t,$ are 
not independent as they share the same eigenvectors.

These findings can be interpreted in the framework of the theory of
free random variables \cite{VDN92}:
if two random matrices $A$ and $B$ of size $N$ are {\sl asymptotically free}
then the average moments of mixed products, e.g.  
$\langle \left| {\rm Tr}A^k (B^{\dagger})^m \right| \rangle,$ 
vanish in the limit $N\to \infty$.
A recent unexpected result of Mingo and Popa  \cite{MP16} 
concerning Haar random unitary matrices states that $U$ and $U^T$ 
are asymptotically free, and in order to substantiate 
our interpretation of the convergence to the Marchenko-Pastur distribution
in Appendix \ref{ApA} we compute numerically 
the averaged traces of products of two random matrices for the cases of interest 
discussed throughout our paper. 
Our results allow us to advance the following

{\bf Conjecture 3.}
{\sl Let $U$ denotes a Haar random unitary matrix of order $N^2$
and $U^R$ the reshuffled matrix. Then for any fixed $t$ the matrices 
$\{U^R, (U^2)^R, \dots, (U^t)^R\}$
are free in the limit $N\to \infty$.} 

This conjecture is supported by recent analytical 
results of Liu et al. \cite{Liu18}, who compute 
the average moments of the Schmidt vector averaged
over the unitary group with respect to the Haar 
measure and show that they converge to Catalan 
numbers characteristic to the Marchenko-Pastur distribution.

\section{Concluding remarks}
\label{sec6}
We have studied the time evolution of a two-qubit unitary gate 
initially in the Cartan form (\ref{cartan}) showing that it 
corresponds to ergodic dynamics inside the $3D$ Weyl chamber,
so that the space average over the set of all non-equivalent orbits
coincides with the average over a single generic trajectory.
Due to a non-intuitive influence of local transformations 
for nonlocality of iterated gates \cite{JMZL17}
this statement does not hold for generic random unitary gates of size four.
However, a stronger property is true in the case of a large system size: 
randomization of nonlocality of bipartite $N \times N$  gates
can be achieved by averaging over a single trajectory $U_0^t$  stemming from a generic 
random gate $U_0$, which asymptotically yields the same results as averaging over the
ensemble of Haar random matrices from $U(N^2)$.

\textbf{Note added.}  After this work was completed 
we learned about the very recent paper \cite{PL18}, 
in which the time dependence of the entangling power 
of the unitary operator describing periodically kicked 
many-body system was used to detect 
non-integrability of the underlying dynamics. 

\begin{acknowledgements} It is a pleasure to thank 
A. Lakshminarayan, J.Mingo, I. Nechita and R. Speicher 
for fruitful discussions.
Financial support by Narodowe Centrum Nauki 
under the grant number DEC-2015/18/A/ST2/00274 
is gratefully acknowledged.
\end{acknowledgements} 
\appendix 

\section{Operator Schmidt Decomposition}
\label{ApA}
A way to study the nonlocality of a quantum gate is to borrow the notion of
the Schmidt decomposition of a vector. 
Any matrix of dimension $N^2$, for instance a unitary gate acting on a bipartite state,
can be treated as a vector in the Hilbert-Schmidt space of matrices
and represented by the \emph{operator Schmidt decomposition,}
\begin{equation} 
\label{genU}
U= N \sum_{i=1}^{N^2} \sqrt{\lambda_i} \; A_i \otimes B_i. 
 \end{equation} 
Here matrices $A_i$  and $B_i$ form  orthonormal bases in the space of operators, 
$\Tr{A_i^\dag A_j}= \Tr{B_i^\dag B_j} = \delta_{ij} $. 
In general the operators  $A_i$  and $B_i$ are not unitary.  
The vector of the expansion coefficients $\vec{\lambda} \equiv (\lambda_1, ..., \lambda_{N^2})$ 
is called \emph{Schmidt vector} and it is invariant with respect to local unitary operations. 

Each matrix $X$ acting on a bipartite  $N \times N$ system can be represented in a 
product basis $\ket{m} \otimes \ket{\mu}$ with matrix element written
$ X_{\stackrel{\scriptstyle m \mu}{n\nu}} = \bra{m \mu}X \ket{n  \nu}.$ 
For any such $X$ one defines the reshuffled matrix $X^R$, 
obtained by reshaping its square blocks of order $N$ into rows of length $N^2$,
with entries 
$ X_{\stackrel{\scriptstyle m \mu}{n\nu}}^R =X_{\stackrel{\scriptstyle m n}{\mu \nu}}$.

Then the vector $\vec{\lambda}(U)$ is equal to eigenvalues of a positive matrix 
\begin{equation}
\label{eq:resh}
\rho = \frac{1}{N^2}U^R (U^R)^\dag
\end{equation}
where $U^R$ stands for the reshuffled matrix \cite{BZ06}. 
\begin{figure}[t!]
\includegraphics[width= 0.98 \columnwidth]{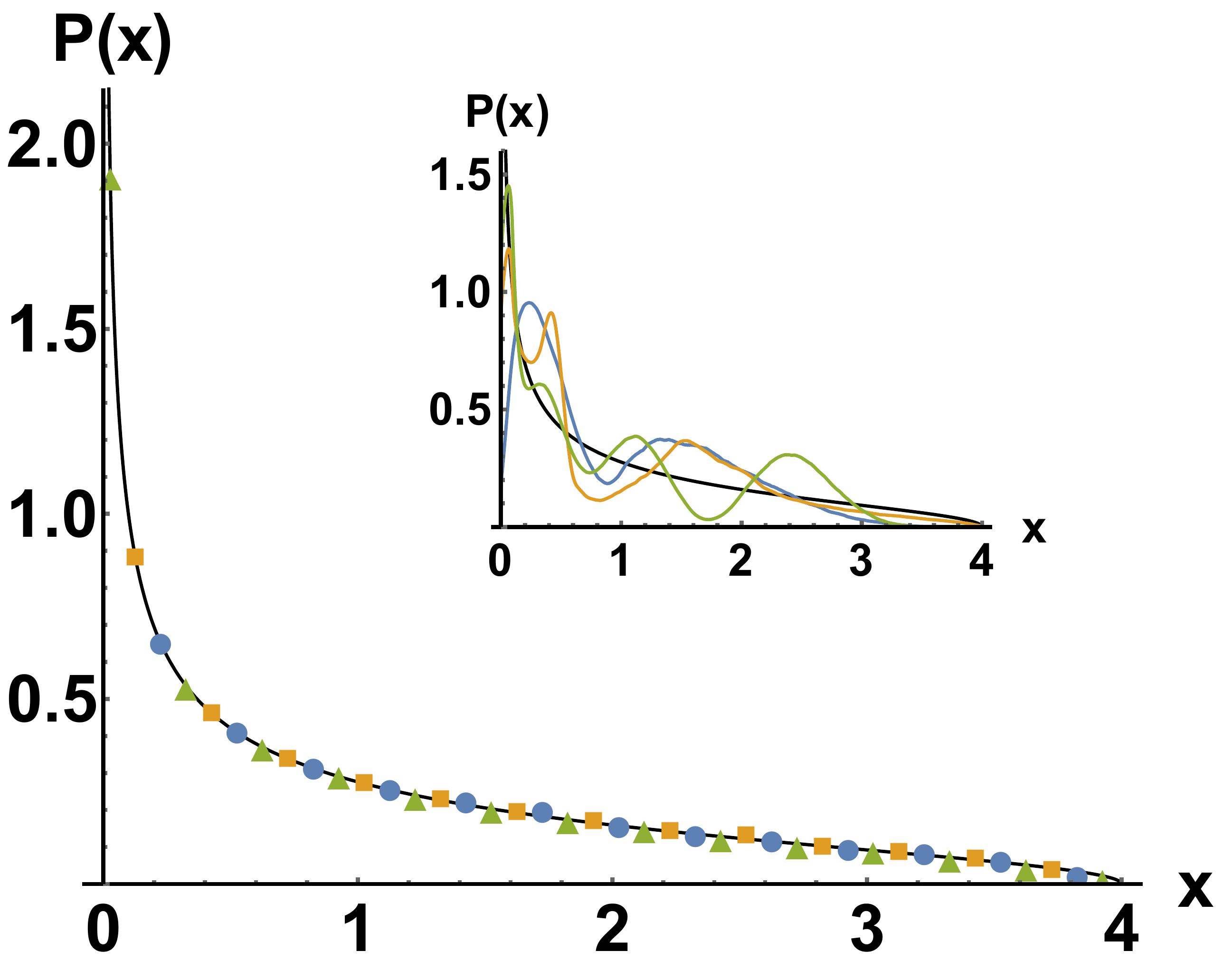}%
\caption{Spectral density P(x) of 3 samples of random
matrices of dimension $d=N^2 = 100$, where $x= d \lambda$ is a rescaled eigenvalue. 
Green triangles refer to Wishart random matrices, 
blue circles to an ensemble of matrices $\rho$ as in (\ref{eq:resh}) 
and yellow squares to an ensemble of non-indipendent matrices 
stemming from a single trajectory $Z^R (Z^R)^\dag/d$, with $Z=U_0^t$. 
Black line denotes Marchenko--Pastur distribution (\ref{MPdistr}).
The differences between analogous distributions
obtained for $d=N^2=4$ with oscillations 
due to level repulsion are shown in inset.
\label{fig:A1} }
\end{figure}
A positive random  matrix $U^R (U^R)^{\dagger}$ has properties of a
Wishart matrix: for a large matrix size the unitarity of $U$ hardly
influences statistical property of the corresponding reshuffled matrix $U^R$,
which can be treated as a typical non-Hermitian matrix from the Ginibre
ensemble \cite{Fo10}, as their statistical properties coincide
asymptotically. 
The last statement is outlined in Fig.\ref{fig:A1}, 
where we show the convergence of the spectral density 
of a random ensemble of states $\rho$ of order $d=N^2$
to the Marchenko--Pastur distribution,
\begin{equation}
\label{MPdistr}
P_{MP}(x) = \frac{\sqrt{1- x/4}}{\pi \sqrt{x}},
\end{equation}
where  $x=d \lambda$ is a rescaled eigenvalue $\lambda$.

In addition we compare $P_{MP}(x)$ with the spectral density of a random ensemble of 
$Z^R (Z^R)^\dag/d$ with $Z = U_0^t$. 
Note that in the limit of $d \gg 1 $
the two distribution coincide, see Fig.\ref{fig:A1}.
  
To analyze possible connections along a trajectory $U^t$
we selected various unitary matrices $U$ according 
to the Haar measure on $U(d)$ and averaged moments 
$\left| {\rm Tr} A^k (B^\dag)^m  \right| $ for 
$ A= U^R$ and $B=(U^t)^R$ and some values of $k$ and $m$. 

Results obtained for two matrices stemming from a single 
trajectory like $U$, $U^2$ are comparable to those 
obtained for two independent random matrices $U_1$ and $U_2$
or two Wishart matrices $W_1$ and $W_2$.  
We report in Tab.\ref{t:freenes} the average values for $m=k=1$. 

\begin{table}[h]
\caption{\label{t:freenes} 
Free independence of two random matrices A  and B of size $d=N^2$ 
assessed by $\mean{\left| {\rm Tr} A B^\dag \right|},$ 
the moment averaged over $10^4$ pairs of matrices:
$W_1$ and $W_2$ denotes two independent Wishart matrices, $U_i$
stands for independent Haar random unitaries, 
while states $\rho_{i}$ are generated from $U_i$ according to Eq.(\ref{eq:resh}).}  
 \vspace{2mm}
\begin{ruledtabular}
\begin{tabular}{c|c|c|c|c} 
&$A$ & $B$ & $d=4$ & $d=100$  \\[1mm] \hline 
&$W_1$ & $W_2$ & $ 0.25 \pm 0.05 $& $ 0.0099 \pm 0.0001  $ \\[.8mm] \hline 
&$\rho_1$ & $\rho_2$ &   $ 0.25 \pm 0.05 $ & $ 0.0100 \pm 0.0001 $ \\[.8mm] \hline 
&$ U_0^R $ & $(U_0^2)^{R}$ & $0.22 \pm 0.11 $ & $ 0.0091 \pm 0.0043 $  \\[.8mm] \hline
&$ U_0^R $ & $(U_0^3)^{R}$ & $ 0.31 \pm 0.16$ & $ 0.0121 \pm  0.0061 $ \\[.8mm] \hline
&$ U $ & $U^T$ & $0.29  \pm 0.20 $& $ 0.0112 \pm 0.0080 $  \\[.8mm] 
\end{tabular}
\end{ruledtabular}
\end{table}
Observe that the averaged traces of products of two random matrices 
decrease with growing dimension $d = N^2,$
and for $d=100$ the differences between 
the mean values are not statistically significant.
Moreover, we have strong numerical evidence  
that reshuffled matrices $(U^t)^R$ along a given trajectory $U^t$
are asymptotically free in analogy to the case $(U, U^T),$
for which a rigorous proof of  freeness 
has been provided \cite{MP16}.  

\section{Time evolution of a two-qubit gate} 
\label{app1}
Aiming to derive relation (\ref{eq:WCdyn}) we start with 
some useful commutation relations for tensor products of Pauli matrices.
An operator, acting on a systems of $n$ qubits, belongs to the group 
$SU(2^n)~\equiv~SU(2)\underset{n-\text{times}}{\otimes ...\otimes}SU(2)$ 
and a particular interesting class of operators for this systems is expressed 
as tensor product of Pauli matrices acting on different qubits. 
With a little bit of algebra it is possible to show that the commutators 
among tensor-products of Pauli matrices are:  
\begin{equation}
\begin{split}
&\left[ \sigma^{\otimes 2m}_i ; \sigma^{\otimes 2m}_j  \right] = 0 , \\
&\left[ \sigma^{\otimes 2m+1}_i ; \sigma^{\otimes 2m+1}_j  \right] = (-1)^m 2 i \epsilon_{ijk} \sigma^{\otimes 2m+1}_k , \\ 
&\left[ \sigma_l \otimes \sigma_j ; \sigma_r \otimes \sigma_s \right] = 
2 i (\epsilon_{lrk} \delta_{js} \sigma_k \otimes \mathds{1}  + \epsilon_{jsk} \delta_{lr} \mathds{1} \otimes \sigma_k ),
\end{split}
\end{equation}
where we assume the standard convention of summation over repeated indexes. 
With use of the first commutator for $m=1$ and the Baker-Campbell-Hausdorff expansion 
any two-qubit gate in the Cartan form,
$V_{(\alpha_1,\alpha_2, \alpha_3)} =
 \exp \left( i \sum_{k=1}^{3} \alpha_k \sigma_k \otimes \sigma_k \right)$,
can be written as a product of three exponential operators, 
\begin{equation}
V =\exp \left( i \sum_{k=1}^{3} \alpha_k \sigma_k \otimes \sigma_k \right) =\prod_{k=1}^{3}\exp  (i  \alpha_k \sigma_k \otimes \sigma_k) .
\end{equation}

The above commutation relations imply the following 
combination law for a product of two two-qubit gates, 
both in the Cartan form,
\begin{equation}
V_{(\alpha_1,\alpha_2, \alpha_3)} V_{(\alpha'_1,\alpha'_2, \alpha'_3)}  = V_{(\alpha_1 + \alpha'_1, \alpha_2 +\alpha'_2 ,\alpha_3 + \alpha'_3)}.
\end{equation}
Considering $t$ actions of a given gate $V$ we get 
$V^t_{(\alpha_1,\alpha_2, \alpha_3)}= V_{(t \alpha_1, t \alpha_2, t \alpha_3)}$
with a restriction that the argument on the right side does not 
need to belong to the Weyl chamber. 
Taking into account reflections from its faces  
we arrive at the required formula for the dynamics 
of a nonlocal gate $$V^t =\exp \{ \vec{\alpha}(V^t) \cdot \vec{\sigma} \},$$ 
where $\vec{\sigma}= (\sigma_1^{\otimes 2},\sigma_2^{\otimes 2},\sigma_3^{\otimes 2})$ 
and the information content of the power of a gate
is given by the ordered vector of the transformed components,
 $\vec{\alpha}(V^t) = 
{\cal {D}}(f (t \alpha_1 ),f (t \alpha_2 ),f (t \alpha_3 ) ).$

\begin{table}[t!]
\caption{\label{t:N2} Mean entropies 
obtained by averaging over $CUE_4$ and $CPE_4$ and 
averages over $m$-dimensional parts  
$\Gamma_m$ of the Weyl chamber.
} \vspace{2mm}
\begin{ruledtabular}
\begin{tabular}{ c | c | c } 
& $\langle S \rangle$ & $\langle S_L \rangle$ \\[1mm] \hline 
$CPE_4$ & $\ln 4  -1 \approx 0.386 $ & $1/4 = 0.25 $ \\[.8mm] \hline 
$1D$ \ set \ $\Gamma_I$ & $ \ln 4 -1   \approx 0.386 $ & $1/4 = 0.25 $ \\[.8mm] \hline 
$2D$ \ set \ $\Gamma_{II}$ & $ 2 (\ln 4  -1) \approx 0.772  $ & $ 7/16 = 0.4375 $ \\[.8mm] \hline
$3D$ \ set \ $\Gamma_{III}$ & $ 1.028 $ & $ 9/16 = 0.5625 $ \\[.8mm] \hline
$CUE_4$ & $ 1.078 $ & $ 3/5= 0.6 $ \\[.8mm]
\end{tabular}
\end{ruledtabular}
\end{table}

\par
\section{Random two-qubit gates}

\label{ApC}

We report in Tab.\ref{t:N2} numerical results of the averages 
for Poissonian ensemble of random diagonal gates and circular 
unitary ensemble of Haar random random unitaries of order 4
and compare them with the analytical results obtained 
by averaging over three distinct subsets $\Gamma_{m}$ of the Weyl Chamber. \par

We emphasize that the mean values of either the entanglement 
and linear entropy over the group $SU(4)$ are  
higher than those derived by averaging over m-dimensional 
subset $\Gamma_m$ of the Weyl chamber. 

The difference between averages over $\Gamma_{III}$ and $CUE_4$
can be explained by local dynamics required to bring 
a generic gate into the Cartan form \cite{JMZL17}.

\end{document}